\documentclass[aip,jcp,reprint]{revtex4-1}

\usepackage{amsmath}
\usepackage{amsfonts}
\usepackage{amssymb}
\usepackage{graphicx}
\usepackage{color}
\usepackage[colorlinks=true,citecolor=red,linkcolor=blue]{hyperref}

\newcommand{\ket}[1]{\left| #1\right\rangle}
\newcommand{\bra}[1]{\left\langle #1\right |}

\begin{document}

\title{Non-Hermitian wave packet approximation for coupled two-level systems \\
 in weak and intense fields}

\author{Raiju Puthumpally-Joseph}
\affiliation{Institut des Sciences Mol\'eculaires d'Orsay (ISMO), CNRS, Univ. Paris-Sud, Universit\'e Paris-Saclay, F-91405 Orsay, France}

\author{Maxim Sukharev}
\affiliation{Science and Mathematics Faculty, College of Letters and Sciences, Arizona State University, Mesa, Arizona 85212, USA}

\author{Eric Charron}
\affiliation{Institut des Sciences Mol\'eculaires d'Orsay (ISMO), CNRS, Univ. Paris-Sud, Universit\'e Paris-Saclay, F-91405 Orsay, France}

\date{\today}

\begin{abstract}
We introduce an accurate non-Hermitian Schr\"odinger-type approximation of Bloch optical equations for two-level systems. This approximation provides a complete description of the excitation, relaxation and decoherence dynamics in both weak and strong laser fields. In this approach, it is sufficient to propagate the wave function of the quantum system instead of the density matrix, providing that relaxation and dephasing are taken into account via automatically-adjusted time-dependent gain and decay rates. The developed formalism is applied to the problem of scattering and absorption of electromagnetic radiation by a thin layer comprised of interacting two-level emitters.
\end{abstract}


\maketitle

\section{Introduction}

Non-Hermitian quantum mechanics (NHQM)\,\cite{PhysRevA.42.10, moiseyev2011non} is an alternative to the standard Hermitian formalism, enabling the solution of otherwise difficult problems. NHQM provides powerful numerical and analytical tools for the study of resonance phenomena\,\cite{PhysRep.302.212, sci.311.1440}. It was proven to be especially useful for a set of problems, for which conventional Hermitian framework fails\,\cite{moiseyev2011non}. NHQM has applications in a variety of fields, including quantum entanglement, cavity quantum electrodynamics, quantum optics when the refractive index is complex, quantum field theory when the parity-time (PT) symmetry properties of the Hamiltonian are investigated\,\cite{Rep.Prog.Phys.70.947}, atomic and molecular physics, and electrical engineering when complex potentials are introduced to simplify numerical calculations\,\cite{PhysRevA.91.062108, EPL.105.40001, PhysRevA.90.054301, PhysRevA.78.062113, PhysRevLett.50.1579,PhysRevLett.105.013903, PhysRevLett.113.250401}.

NHQM finds important application in studies of the dynamics of open systems\,\cite{PhysRevA.85.032111}. Open systems, which are coupled to one or more continua have dissipative processes determined by couplings the reservoir. Treating the dissipation dynamics is often complicated. Non-Hermitian Hamiltonians can be taken as an effective tool for treating such processes. To achieve a proper description of the system dynamics using non-Hermitian Hamiltonians, the real parts of the eigenvalues are related to the eigenenergies and the imaginary parts can be linked to decay rates. A non-Hermitian model was presented in\,\cite{jcp.138.024108} as an alternative to the well-known Bloch equations for treating the dynamics of a collection of two-level and multi-level quantum systems in the weak field regime. This model works well irrespective of the strength of the emitter-emitter interaction but it is limited to weak excitations. It is, for example, able to describe collective effects in dense atomic vapors\,\cite{PhysRevLett.113.163603, PhysRevA.91.043835}. However the model fails to describe the dynamics of the system in the case of a strong excitation induced, for instance, in the presence of intense laser fields. In this paper, we propose to extend this model to the strong excitation regime. We show that the extended model can accurately describe the dissipative dynamics of a collection of interacting two-level systems excited by a strong field.

The simplest quantum system, a two-level system, is widely used to model physical problems which are of great interest\,\cite{RevModPhys.73.565}. Spin states of electrons are the best example of naturally occurring two-level systems\,\cite{RevModPhys.55.855, RevModPhys.79.1217}. Recent advances in laser technology provide a high precision control over electromagnetic field parameters that enable to tune the exciting fields in resonance with specific quantum levels. In such a case, the quantum dynamics can often be treated within the framework of a two-level system. It is thus widely used to understand the quantum dynamics in various systems\,\cite{PhysRevB.32.4410, PhysRevLett.42.974, PhysRevLett.67.516}. Two-level systems are also investigated widely for manipulating laser-matter interaction which can lead to technological breakthroughs\,\cite{PhysRevLett.104.083901, Pirkkalainen2015, Lisenfeld2015, Zrenner2002}. The model presented in this paper is applied to two-level systems, both in the weak and strong interaction regimes.

The paper is organized as follows. Current theoretical models are briefly discussed and compared with our new approach in section\;\ref{sec:tbm}. Our model is applied to a collection of two-level systems and the studies are carried out in both the weak and intense field regimes. The weak probe regime is discussed in section\;\ref{wpr} and the results obtained for intense fields are discussed in section\;\ref{ipr}. The work is summarized in section\;\ref{sec:summary}.

\section{Theoretical background}
\label{sec:tbm}

\subsection{Bloch equations}
\label{sec:tls}

We consider a two-level quantum system consisting of levels $\ket{1}$ and $\ket{2}$,
with eigenfrequencies $\omega_1$ and $\omega_2$, respectively. The system is subject to an electromagnetic field whose career frequency $\omega_0$ is close to the transition frequency $\omega_{B}=\omega_2-\omega_1$. The time dynamics of such a system is represented by the dissipative Liouville-von Neumann equation for the density matrix $\hat\rho(t)$\,\cite{Blum:1996kn} 
\begin{equation}
 \label{eq:Liouville}
 i\hbar\,\partial_t\hat\rho = [ \hat H , \hat \rho ] -i\hbar \hat\Gamma \hat\rho\,,
\end{equation}
where $\hat H = \hat H_0 + \hat V_i(t)$ is the total Hamiltonian and $\hat\Gamma$ is the relaxation super-operator taken in the Lindblad form.  The non-diagonal elements of $\hat\Gamma$ include a pure dephasing rate $\gamma^*$, and the diagonal elements of this operator consist of the radiationless decay rate $\Gamma$ of the excited state under Markov approximation\,\cite{Breuer02, schlosshauer2007decoherence}.  The field free Hamiltonian $ \hat H_0 $ of the system can be written in terms of the diagonal elements of the density matrix as
\begin{equation}
 \label{eq:H0}
 \hat H_0 = \hbar\omega_1 \ket{1}\!\!\bra{1} + \hbar\omega_2 \ket{2}\!\!\bra{2}.
\end{equation}
The interaction of the system with the applied field is written in the dipole approximation
\begin{equation}
 \label{eq:Vi}
 \hat V_i(t) = \hbar\Omega(t) \big( \ket{2}\!\!\bra{1} + \ket{1}\!\!\bra{2} \big),
\end{equation}
where $\Omega(t)$ represents the instantaneous Rabi frequency describing the coupling between the quantum system and the applied field. Eqs.\,(\ref{eq:Liouville})-(\ref{eq:Vi}) lead to a set of first order differential equations describing the dynamics of two-level quantum systems usually referred to as optical Bloch equations\,\cite{Allen:1975ij}
\begin{subequations}
 \begin{eqnarray}
  \partial_t\rho_{11} & = & i\Omega(t)(\rho_{12}-\rho_{21})+\Gamma\rho_{22}\,,\label{rho11}\\
  \partial_t\rho_{12} & = & i\Omega(t)(\rho_{11}-\rho_{22})+(i\omega_B-\gamma)\rho_{12}\,,\label{rho12}\\
  \partial_t\rho_{21} & = & i\Omega(t)(\rho_{22}-\rho_{11})-(i\omega_B+\gamma)\rho_{21}\,,\label{rho21}\\
  \partial_t\rho_{22} & = & i\Omega(t)(\rho_{21}-\rho_{12})-\Gamma\rho_{22}\,,\label{rho22}
 \end{eqnarray}
 \label{rho}
\end{subequations}
where $\gamma=\gamma^*+\Gamma/2$.

\subsection{Non-Hermitian wave packet approximations}
\label{sec:nhtlwpa}

The dynamics of two-level quantum systems can also be represented using the non-Hermitian formalism. In this case the quantum state of the two-level system is represented as a wave packet $\ket{\Psi(t)}$ formed by the superposition of the two states $\ket{1}$ and $\ket{2}$ with time-dependent coefficients $c_1(t)$ and $c_2(t)$ as
\begin{equation}
 \ket{\Psi(t)} = c_1(t) \ket{1} + c_2(t) \ket{2}. \label{expansion}
\end{equation}
In the usual case, the wave packet given in Eq.\,(\ref{expansion}) is the solution of the time-dependent Schr\"odinger equation (TDSE) described by the Hamiltonian $ \hat H $ which is a Hermitian operator. The addition of empirical imaginary parts $+i\hbar\gamma_1(t)/2$ and $-i\hbar\gamma_2(t)/2$ to the eigenenergies of the system leads to a non-Hermitian dynamics. $\gamma_1(t)$ is the gain factor of the ground state and $\gamma_2(t)$ is the decay rate of the excited state.

It can be shown by projecting the TDSE onto the states $\ket{1}$ and $\ket{2}$ that the new coefficients $c_1(t)$ and $c_2(t)$ obey a set of coupled differential equations\,\cite{jcp.138.024108}
\begin{equation}
i\partial_t\begin{pmatrix}c_1\\c_2\end{pmatrix} = 
\begin{pmatrix}\omega_1 + i\gamma_1(t)/2 & \Omega(t)\\\Omega(t) & \omega_2 - i\gamma_2(t)/2 \end{pmatrix}
\begin{pmatrix}c_1\\c_2\end{pmatrix}. \label{c1c2}
\end{equation}
Taking $\rho^{s}_{ij}(t)=c_i(t)\,c_j^*(t)$ we can write equations for the modified density matrix elements $\rho^{s}_{ij}(t)$, where the superscript $s$ represents the non-Hermitian Schr\"odinger-type model
\begin{subequations}
 \begin{eqnarray}
 \partial_t\rho^s_{11} & = & i\Omega(t)(\rho^s_{12}-\rho^s_{21})+\gamma_1\,\rho^s_{11}\,,\label{rhos11}\\
 \partial_t\rho^s_{12} & = & i\Omega(t)(\rho^s_{11}-\rho^s_{22})+\big[i\omega_B-\frac{\gamma_2-\gamma_1}{2}\big]\rho^s_{12}\,,\label{rhos12}\\
 \partial_t\rho^s_{21} & = & i\Omega(t)(\rho^s_{22}-\rho^s_{11})-\big[i\omega_B+\frac{\gamma_2-\gamma_1}{2}\big]\rho^s_{21}\,,\label{rhos21}\\
 \partial_t\rho^s_{22} & = & i\Omega(t)(\rho^s_{21}-\rho^s_{12})-\gamma_2\,\rho^s_{22}\,.\label{rhos22}
\end{eqnarray}
\end{subequations}
The empirical gain and decay factors $\gamma_1(t)$ and $\gamma_2(t)$ are the parameters that have to be modeled properly so that the dynamics of the system is reproduced accurately.

Using the first-order perturbation theory we can accurately describe the coherences of the system\,\cite{jcp.138.024108}. By comparing Eqs.\,(\ref{rhos12}) and\;(\ref{rhos21}) with Eqs.\,(\ref{rho12}) and\;(\ref{rho21}), the empirical gain and decay factors can be related to the decay and decoherence rates of the system via
\begin{equation}
 \gamma_2(t)-\gamma_1(t) = 2\gamma = 2\gamma^*+\Gamma\,. \label{coh}
\end{equation}
In the non-Hermitian model presented in Ref. [\onlinecite{jcp.138.024108}], which we call non-Hermitian model number 1 (NH1), an additional condition on the conservation of the total norm \emph{i.e},
\begin{equation}
\rho^s_{11}(t)+\rho^s_{22}(t) = 1 \label{norm}
\end{equation}
was taken into account. This yields the following definition of the empirical gain and decay factors 
\begin{subequations}
 \begin{eqnarray}
  \gamma_1(t) & = & \frac{2\gamma\,|c_2(t)|^2}{|c_1(t)|^2-|c_2(t)|^2}\,, \label{g1}\\[0.15cm]
  \gamma_2(t) & = & \frac{2\gamma\,|c_1(t)|^2}{|c_1(t)|^2-|c_2(t)|^2}\,. \label{g2}
 \end{eqnarray}
 \label{gam}
\end{subequations}
These rates can be plugged into Eq.\,(\ref{c1c2}), which can be solved to find the complex coefficients $c_1(t)$ and $c_2(t)$ and hence to describe the dynamics of the two-level system accurately. The details and applications of this approach can found in Ref. [\onlinecite{jcp.138.024108}].

NH1 model fails when the system is exited with a high intensity pulse. Indeed, it is evident from Eqs.\,(\ref{g1}) and\;(\ref{g2}) that the empirical gain and decay factors chosen to mimic the dynamics of the system diverge when the exited state is populated at $50\% $. Thus it is clear that NH1 is inadequate to describe the dynamics of two-level systems driven by an intense laser field. It is the conservation of the norm, \emph{i.e} Eq.\,(\ref{norm}), which brings this pole to the model. We show below that if we drop this condition it is possible to solve this problem and to describe the coherence dynamics accurately.

From the dynamical Eqs.\,(\ref{rho12}) and\;(\ref{rho21}), it is clear that an accurate description of the coherences $\rho_{12}(t)$ and $\rho_{21}(t)$ requires an accurate evaluation of the population difference $\Delta(t)=\rho_{22}(t)-\rho_{11}(t)$. From Eqs.\,(\ref{rho22}) and\;(\ref{rho11}) we obtain
\begin{equation}
 \partial_t\Delta = 2i\Omega(t)(\rho_{21}-\rho_{12})-2\Gamma\rho_{22}\,.
 \label{delta}
\end{equation}
A similar evolution equation can be derived from the non-Hermitian model by taking the difference between Eqs.\,(\ref{rhos22}) and\;(\ref{rhos11})
\begin{equation}
\partial_t\Delta^{\!s} = 2i\Omega(t)(\rho^s_{21}-\rho^s_{12})-(\gamma_1\,\rho^s_{11}+\gamma_2\,\rho^s_{22})\,.
\label{deltas}
\end{equation}
By comparing Eq.\,(\ref{deltas}) with Eq.\,(\ref{delta}) we see that a correct description of the population difference requires that
\begin{equation}
 \gamma_{1}(t)\,|c_1(t)|^2 + \gamma_{2}(t)\,|c_2(t)|^2 = 2\Gamma\,|c_2(t)|^2\,.
 \label{Eq:Gamma-condition-2}
\end{equation}
And from Eqs.\,(\ref{Eq:Gamma-condition-2}) and\;(\ref{coh}) we finally get
\begin{subequations}
  \begin{eqnarray}
  \gamma_1(t) & = & \frac{(\Gamma-2\gamma^*)\,|c_2(t)|^2}{|c_1(t)|^2+|c_2(t)|^2}\,,\label{g1s}\\[0.15cm]
  \gamma_2(t) & = & \frac{2\gamma\,|c_1(t)|^2+2\Gamma\,|c_2(t)|^2}{|c_1(t)|^2+|c_2(t)|^2}\,.\label{g2s}
  \end{eqnarray}
\end{subequations}
The time-dependent gain and decay rates derived above is the main result of the manuscript. To compare it with other approaches it will be called NH2 onwards.

It is clear that Eqs.\,(\ref{g1s}) and\;(\ref{g2s}) do not have any pole compared to Eqs.\,(\ref{g1}) and\;(\ref{g2}). Thus these refined gain and decay factors can be used for simulating the system dynamics in both weak and strong fields. An application and a comparison of these models are discussed in the next section.

\section{Application to a uniform nano-layer}
\label{sec:aunla}

\begin{figure}[ht]
\begin{center}
\includegraphics[width=8.5cm]{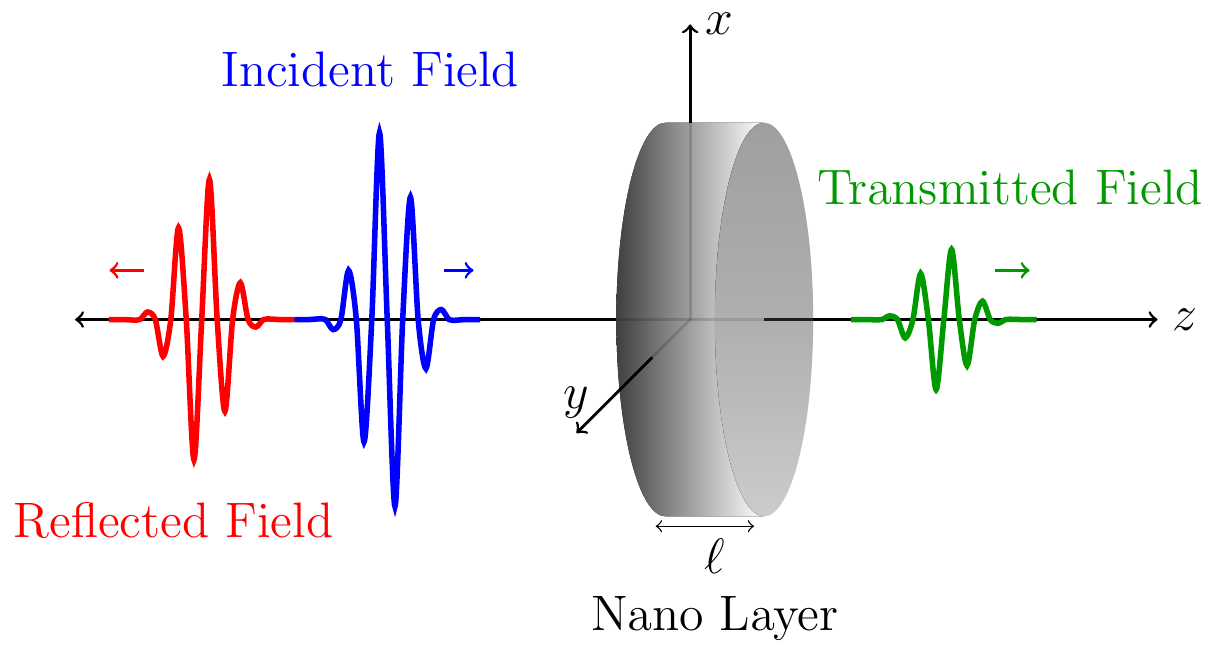}
\caption{(Color online) A schematic view of the system. An infinite slab of width $\ell$ composed of identical two-level emitters is exposed to a laser pulse. A part of the EM field is reflected and the remaining part is either transmitted through or absorbed by the layer of emitters.}
\label{fig_system}
\end{center}
\end{figure}

In this section we consider a collection of identical two-level systems assumed to form a uniform layer consisting of $n$ emitters per cm$^{3}$. It is also assumed that this layer is finite in $z$ direction and infinite in both $x$ and $y$. All emitters are prepared initially in their ground state $\ket{1}$. An $x$-polarized incident electromagnetic (EM) pulse of duration $\tau$ propagating in the $z$-direction interacts the layer, as shown in Fig.\,\ref{fig_system}. A part of the incident radiation is reflected and the remaining part passes through the system. The EM field propagating in the medium is also partially absorbed by the coupled two-level emitters, setting up the dynamics and other associated effects as described, for example, in Ref. [\onlinecite{PhysRevLett.113.163603, PhysRevA.91.043835, PhysRevA.61.063815, PhysRevA.91.043810, PhysRevA.84.053428}].

The dynamics of the EM field $E_x(z,t)$ and $H_y(z,t)$ is described by Maxwell's equations in the time domain
\begin{subequations}
 \begin{eqnarray}
  \mu_0\,\partial_t H_y      & = & -\partial_z E_x, \label{Faraday} \\
  \epsilon_0\,\partial_t E_x & = & -\partial_z H_y -\partial_t P_x, \label{Ampere}
 \end{eqnarray}
\end{subequations}
where $\mu_0$ and $\epsilon_0$ are the permeability and permittivity of the free space, respectively. These equations are implemented and solved using a generalized finite-difference time-domain technique where both the electric and magnetic fields are propagated in discretized spatial and temporal grids as described in Ref. [\onlinecite{taflove2005computational}]. The interaction of the EM fields with the two-level emitters polarizes the medium giving rise to a macroscopic polarization $P_x(z,t)$ which is related to the microscopic transition dipoles $\mu_x$ via
\begin{equation}
 P_x(z,t)= n\,\langle \mu_x \rangle, \label{polarization}
\end{equation}
where $\langle \mu_x \rangle$ stands for the expectation value of the point-wise dipole at position $z$ and time $t$. The polarization of the medium results in a modification of the effective local electric field. This modification is taken into account for the  proper inclusion of the dipole-dipole couplings in the system. This is especially important when the system density increases. It can be done by defining the local field $E_{\text{loc}}(z,t)$ experienced by the emitters using the well-known Lorentz-Lorenz correction\,\cite{jackson1962classical},
\begin{equation}
\label{LL}
E_{\text{loc}}(z,t) = E_x(z,t) + \frac{P_x(z,t)}{3\epsilon_0}\,.
\end{equation}

The polarization $P_x(z,t)$ of Eq.\,(\ref{polarization}) is the link between the field dynamics described by Eqs.\,(\ref{Faraday}) and\;(\ref{Ampere}) and the quantum description of the system\,\cite{Judkins:95,3167407, Fratalocchi:2008aa, Sukharev:2011ls, jcp.138.024108}. The system is  modeled using both the exact coupled Maxwell-Liouville dynamical equations (\ref{rho}) and the non-Hermitian equations (\ref{c1c2}) in their NH1 and NH2 variations. The density matrix elements $\rho_{ij}(z,t)$ and the wave packet components $c_i(z,t)$ are calculated using the fourth order Runge-Kutta method. The calculated transmitted and reflected fields give access to the corresponding Poynting vector
\begin{equation}
 S(\omega) = \left| \widetilde{E}_x(\omega) \, \widetilde{H}_y(\omega) \right|,
\end{equation}
where $\widetilde{E}_x(\omega)$ and $\widetilde{H}_y(\omega)$ are the Fourier components of the transmitted or reflected EM fields. Their evaluation and subsequent normalization with respect to the incident energy flux allows to calculate the reflection, $R(\omega)$, and transmission, $T(\omega)$. The part of the energy absorbed and dissipated by the medium can be caclulated as $A(\omega)=1-T(\omega)-R(\omega)$.

The different models describing the quantum dynamics of two-level systems in the medium can be compared in the weak as well as in the strong excitation regime. To compare the new model NH2 with the well-established Maxwell-Liouville model and the recently introduced non-hermitian approach NH1, we calculate transmission and reflection from a layer of thickness $\ell = 600$\,nm. The medium is composed of identical two-level emitters with transition energy $\hbar\omega_B = 2$\,eV and transition dipole $\mu_x = 4.0$\,D. The dephasing rate is $\gamma^* = 10$\,THz and the nonradiative decay rate is $\Gamma = 1$\,THz. The layer is excited with a laser pulse with a Gaussian envelop of duration $\tau = 10$\,fs (FWHM) whose career frequency is resonant with the transition frequency. The response of the system towards the incident field is finally calculated as a function of the relative detuning $\delta = (\omega - \omega_B)/\gamma$.

Media composed of a large number of interacting dipoles can respond to the incident EM field in a collective manner due to the strong dipole-dipole interactions. As demonstrated in Ref. [\onlinecite{PhysRevA.91.043835}] for the same kind of geometry in the linear excitation regime, the strength of the dipole-dipole interaction can be quantified by a dimensionless parameter $\eta = \Delta/\gamma$, where $\Delta = n\mu_x^2/(9\hbar\epsilon_0)$. Since the cooperative nature of the system can significantly alter its dynamics, the calculations are done in both the weak and strong interaction regimes. The cooperative behavior of the system is setting up gradually as the density increases. It induces a considerable modification in the local fields. The validity of the proposed model NH2 has to be checked in both limits. We consider two specific cases of interaction regimes: one with $\eta = 1.3 \times 10^{-7}$ which shows no cooperative response and the second case with $\eta = 1.3$ exhibiting a strong cooperative behavior.

\subsection{Weak excitation limit}
\label{wpr}

In the weak probe regime, the system is excited with a laser pulse of small amplitude $E_0$. For the numerical simulations we have chosen $E_0 = 1$\,V/m, such that the system responds linearly to the field with an excitation probability much smaller than one. If the density is small, the atoms respond almost independently. On the contrary, in the strong interaction regime reached at large densities, dipole-dipole interactions alter the response of the system, resulting in a dramatic change of reflection and transmission spectra.

\begin{figure}[ht]
\begin{center}
\includegraphics[width=8.5cm]{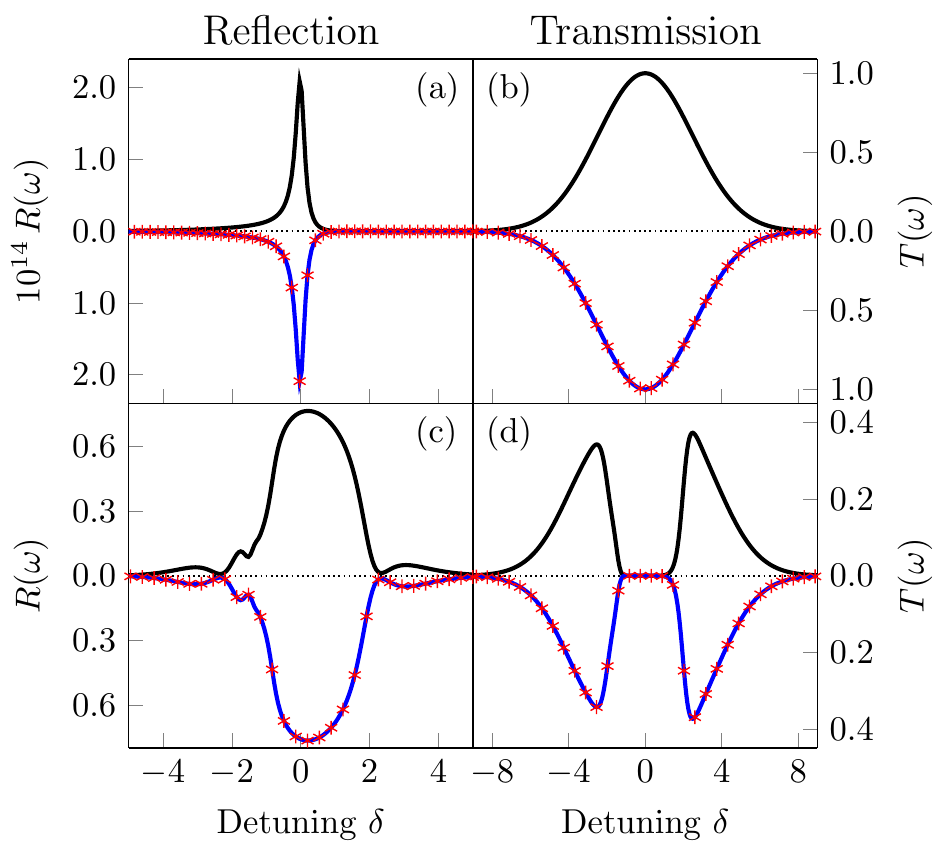}
\caption{(Color online) Response of the medium towards a weak probe. The black curves are the responses calculated using the Maxwell-Liouville model. The responses calculated via the non-Hermitian models are shown as inverted plots. The blue curves and the red asterisks represent NH1 and NH2 models respectively. Panels (a) and (b) show the reflection and transmission from the layer consisting of weakly interacting two-level systems, with $\eta = 1.3 \times 10^{-7}$. Panels (b) and (c) are the same for strongly interacting two-level systems, with $\eta = 1.3$.}
\label{fig_weak_field}
\end{center}
\end{figure}

Fig.\,\ref{fig_weak_field} shows the response calculated for a weak probe. The results obtained using the quantum dynamics of the system via optical Bloch equations are shown as black curves. The results for NH1 and NH2 models are shown as inverted plots as they are on top of the black curve. The blue curves are obtained by modeling the system via NH1 and the  red marks via the new model NH2.

Panels (a) and (b) show the reflection and transmission from the nano-layer of weakly interacting systems. The transmitted pulse is almost identical to the incident pulse due to the weak reflection and weak excitation of the system. One can notice that the three models agree perfectly irrespective of the different descriptions of coherences that are taken into account, even in the strong interaction regime which is shown in panels (c) and (d). In this regime, the system responds like a dissipative mirror near the transition energy even though it is weakly excited\,\cite{PhysRevLett.113.163603, PhysRevA.91.043835}. The non-Hermitian models NH1 and NH2 follow all the features of this cooperative behavior perfectly, proving their ability to describe the dynamics of two-level coupled systems flawlessly within the limits of the approximation taken into account.

\begin{figure}[ht]
\begin{center}
\includegraphics[width=8.5cm]{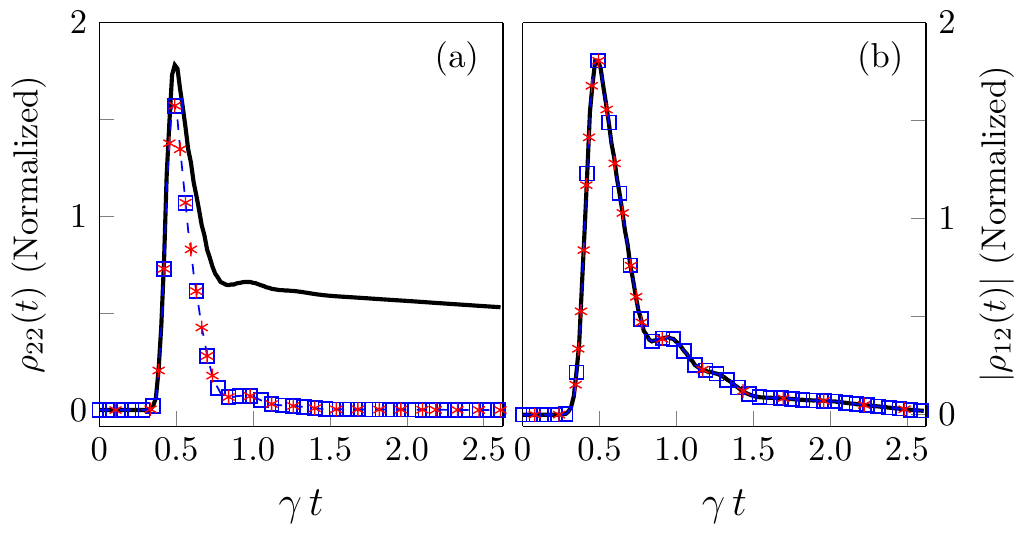}
\caption{(Color online) Panel (a) shows the excited state population $\rho_{22}(t)$ and panel (b) shows the absolute value of the coherence $|\rho_{12}(t)|$ in the strong interaction regime corresponding to $\eta=1.3$ calculated $ 290 $ nm inside the nano-layer. The black curves show the results obtained via the Maxwell-Liouville model. The dashed blue curve with squares and the red asterisks depict the results of NH1 and NH2 models respectively. All other parameters are as in Fig.\,\ref{fig_weak_field}.}
\label{fig_pop_coh}
\end{center}
\end{figure}

In the non-Hermitian approximations NH1 and NH2, the accuracy obtained in describing the optical response of the system is achieved because the coherence $\rho_{12}(t)$ is correctly described due to the choice made in Eq.\,(\ref{coh}). This is however obtained at the cost of loosing the accuracy in the description of the excited state population $\rho_{22}(t)$, whose decay rate differs significantly from its expected value $\Gamma$ since it equals to $\gamma_2(t)$ in this case.

Fig.\,\ref{fig_pop_coh} compares the evolution of the excited state populations and coherences calculated at a point $ 290 $ nm inside the nano-layer using two different models. As previously, the results obtained with the optical Bloch equations are shown in black. The blue curves with squares show the NH1 results and the red asterisks are for NH2 model. Initially, the excited state populations shown in panel (a) and described by the two non-Hermitian models follow the \textit{exact} dynamics but they decay far too quickly since in this particular case $\gamma_2(t)$ quickly exceeds $\Gamma$. On the contrary, the evolution of the coherences shown in panel (b) are identical in the different models considered. This ensure a correct description of the system's optical response since the macroscopic polarization is proportional to the coherence in the linear regime.

It is also worth noticing in panels (a) and (b) of Fig.\,\ref{fig_pop_coh} that the excited state populations and the coherences of NH1 and NH2 models show the same evolution. This is due to the fact that in the weak excitation limit, \emph{i.e} $|c_1(t)|^2 \gg |c_2(t)|^2$, the decay rates $\gamma_2(t)$ of these two models are identical.

\subsection{Strong excitation limit}
\label{ipr}

It was already mentioned that NH1 model is expected to fail as the intensity of the applied field increases. This is due to the presence of a pole in the definition\;(\ref{gam}) of the gain and decay rates $\gamma_1(t)$ and $\gamma_2(t)$ when $|c_1(t)|^2 = |c_2(t)|^2$. The improved model NH2 is designed to avoid this unphysical behavior.

\begin{figure}[ht]
 \begin{center}
  \includegraphics[width=8.5cm]{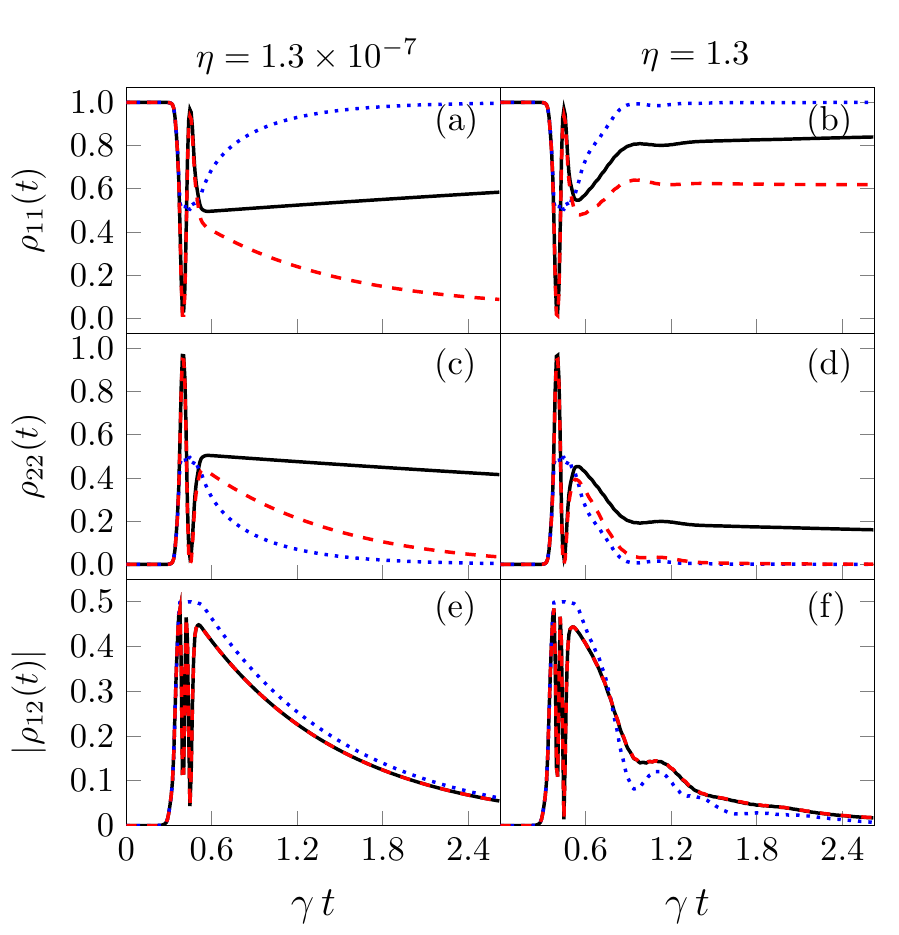}
  \caption{(Color online) Evolution of the populations and coherences of the system in the strong excitation regime. The black curves show the evolution of the density matrix elements using the Maxwell-Liouville model as functions of time in units of $1/\gamma$. The dotted blue and dashed red curves represent the evolution of the density matrix elements of NH1 and NH2 models respectively. Panels (a), (c) and (e), on the left, are for a layer consisting of weakly interacting two-level emitters with $\eta = 1.3 \times 10^{-7}$. Panels (b), (d) and (f), on the right, are for strongly interacting two-level emitters with $\eta = 1.3$. The values are calculated at a point $ 290 $ nm inside the cell.}
  \label{fig_SF_dens_mat}
 \end{center}
\end{figure}

To test the validity of NH2 compared to optical Bloch equations, the system is now excited by a high incident field with a peak amplitude of $E_0 = 10^{10}$\,V/m (about 0.02\,au). This field is high enough to significantly populate the excited state. Fig.\,\ref{fig_SF_dens_mat} shows the evolution of the density matrix elements describing the quantum dynamics of the system for three different models.

Panels (a) and (b) are the ground state populations in the weak and strong interaction regimes respectively. Similarly, panels (c) and (d) show the excited state populations and panels (e) and (f) show the coherences. The reference populations and coherences obtained by integrating optical Bloch equations are shown as solid black curves. As expected, the NH1 results (blue dotted curves) deviate significantly at early times from the optical Bloch model, even for the coherences. It can be noticed that this deviation appears as soon as the system approaches an inversion of population, with $|c_1(t)|^2 = |c_2(t)|^2$.

When the excited state population reaches almost 50\% it exhibits fast oscillations proving clearly the inability of NH1 model to attain higher excited state population due to a very fast increase (divergent pole) of the empirical gain and decay factors $\gamma_1(t)$ and $\gamma_2(t)$. Since the populations $|c_1(t)|^2$ and $|c_2(t)|^2$ are used to evaluate $\gamma_1(t)$ and $\gamma_2(t)$ it affects the accuracy of the coherences in the system, as can be seen in panels (e) and (f). In both the weak and strong interaction regimes, NH1 model is inadequate for the description of the dynamics initiated by strong fields.

On the other hand, as seen in panels (e) and (f), the new model NH2 (dashed red line) follows perfectly the coherence of the system in both weak and strong interaction regimes in intense fields. This is obtained at the cost of losing the accuracy in the description of the populations, as we can see in panels (a), (b), (c) and (d).

\begin{figure}[ht]
\begin{center}
\includegraphics[width=8.5cm]{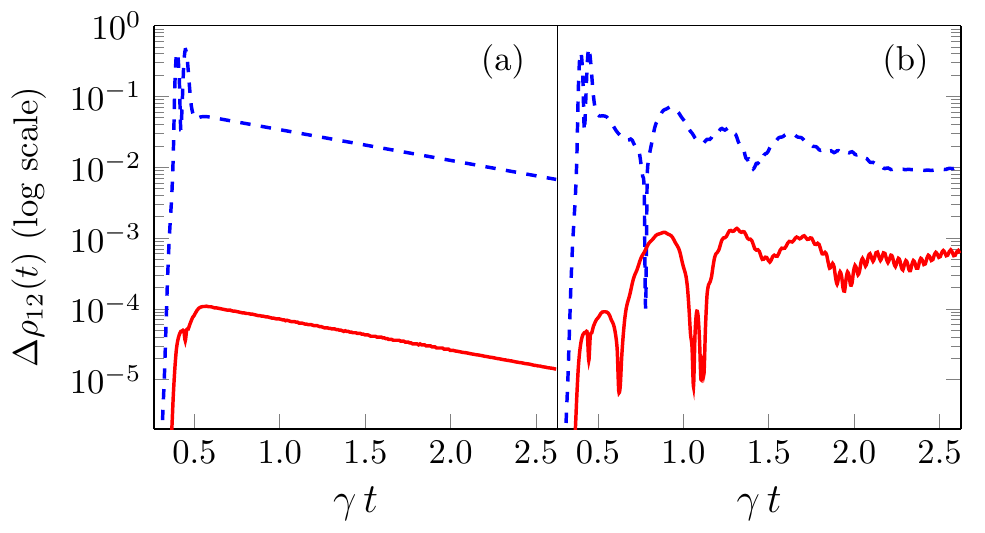}
\caption{(Color online) Differences in coherences $\Delta\rho_{12}(t)$ in strong field dynamics. The dashed blue curves are the absolute value of the difference between the coherences obtained with the optical Bloch equations and with NH1 model. The red curve is the same difference but with respect to NH2 model. Panel (a) (log scale) is the weak interaction regime corresponding to $\eta = 1.3 \times 10^{-7}$ and the panel (b) (log scale) is the strong interaction regime with $\eta = 1.3$.}
\label{fig_error_coh}
\end{center}
\end{figure}

To see how accurately the non-Hermitian models mimic the coherence dynamics of two-level systems in strong fields, the difference $\Delta\rho_{12}(t)=|\rho_{12}(t)-\rho_{12}^s(t)|$ of the coherence calculated using the optical Bloch equations and using the non-Hermitian models are shown in Fig.\,\ref{fig_error_coh} in a semi-logarithmic scale.

\begin{figure}[ht]
 \begin{center}
  \includegraphics[width=8.5cm]{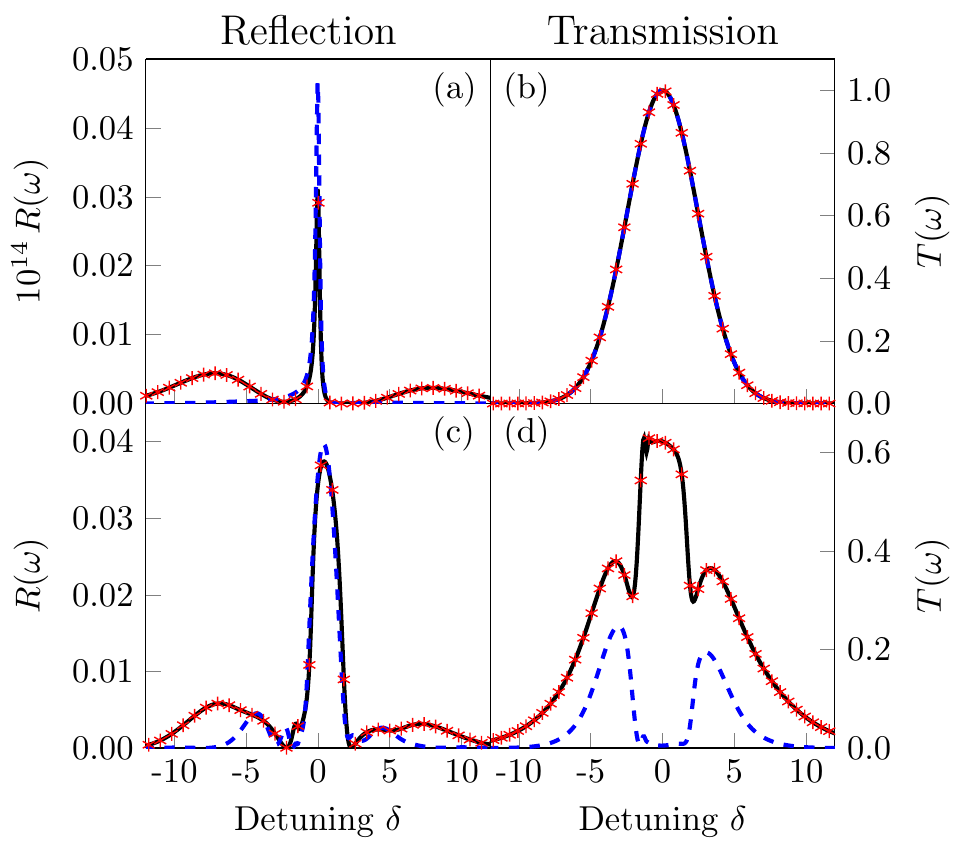}
  \caption{(Color online) System's response to a strong probe. The nano-layer is excited with a  Gaussian pulse of FWHM $ 10 $ fs and peak amplitude $ 0.02 $ au. The black curves are the responses calculated using the Maxwell-Liouville model. The dashed blue curve and the red asterisks represent the results obtained with NH1 and NH2 models respectively. Panels (a) and (b) show the reflection and transmission from the layer consisting of weakly interacting two-level systems with $\eta = 1.3 \times 10^{-7}$. Panels (c) and (d) show the same spectra for strongly interacting two-level systems with $\eta = 1.3$.}
  \label{fig_strong_field}
 \end{center}
\end{figure}

The dashed blue curve shows $\Delta\rho_{12}(t)$ as a function of time in units of $1/\gamma$ for NH1 model and the red curve is for NH2 model. Panel (a) shows the comparison in the weak interaction regime and panel (b) shows the comparison in the strong interaction regime. It is clear that the introduction of NH2 model improves greatly the error, which is always smaller than 1\%, in comparison with the previous model whose accuracy is not acceptable in strong fields. Finally, the calculated optical response of the system is shown in Fig.\,\ref{fig_strong_field}.

Panels (a) and (b) are the reflected and transmitted pulses from the weakly interacting system while panels (c) and (d) are the same for strongly interacting systems. The black curves are the results of the Maxwell-Liouville model. The red marks, which are the results of NH2 model, are always in very good agreement, thus validating the approximations made to derive the empirical gain and decays factors in NH2 model.

In the weak interaction regime, the system transmits almost all the incident energy. This is the reason for the agreement of the transmitted pulse obtained with NH1 model (dashed blue curve) in panel (b). However, for the reflection seen in panel (a), even if it is small, NH1 model shows a clear mismatch with the optical Bloch equations, not only in the peak value, but also on the two sidebands which appear due to the strong excitation of the system. On the contrary, NH2 model (red asterisks) is in perfect agreement with the optical Bloch equations for all values of the detuning.

The failure of NH1 model, and in contrast the striking validity of NH2 model, are even more pronounced when the dipoles interact strongly, as seen in panels (c) and (d). Along with the nonlinearities introduced by the intense field, the dipole-dipole interactions modify the system response to a large extend\,\cite{PhysRevLett.113.163603}. All features of the spectra obtained using the Bloch optical model are very well reproduced by NH2 model, proving its ability to deal with any interaction regime. In panel (c) we see that the transmission is clearly affected by the collective response of the medium\,\cite{PhysRevLett.113.163603}, a feature that the first model fails to reproduce.

\section{Summary and conclusions}
\label{sec:summary}

In this paper, we have proposed an improved non-Hermitian approximation of the optical Bloch equations where one propagates the wave function of a quantum system instead of a complete density matrix. This method provides an accurate description of the dynamics of single two-level emitters, as well as ensembles of coupled two-level emitters in both the weak and strong excitation limits, thus allowing numerical simulations in strong fields. In the case of coupled two-level emitters, the proposed model takes into account the collective optical response of coupled emitters.

\section*{Acknowledgements}

R.P.J. and E.C. acknowledge support from  the EU (Project ITN-2010-264951, CORINF). M.S. is grateful to the Universit\'e Paris-Sud (Orsay) for financial supports through an invited Professor position in 2016. M.S. also would like to acknowledge financial support by AFOSR under grant No. FA9550-15-1-0189.  We also acknowledge the use of computing cluster GMPCS of the LUMAT federation (FR LUMAT 2764).

\bibliographystyle{apsrev4-1}

\end{document}